\documentclass[12pt,osajnl,preprint,showpacs]{revtex4}
\DeclareRobustCommand{\baselinestretch{2.2}}

\usepackage{graphicx}
\usepackage{xspace}
\begin{document}
\newcommand{\vref}{\ref}


\newcommand\affilmops{\affiliation{Laboratoire Mat\'eriaux Optiques, Photonique et
Syst\`{e}mes\\ Unit\'e de Recherche Commune \`{a} l'Universit\'e de Metz et
\`{a} Sup\'elec\\ CNRS-UMR 7132
\\ 2 rue Edouard Belin, 57070 Metz, France}}

\newcommand\affillnio{\affiliation{Laboratoire de Nanotechnologie et d'Instrumentation Optique\\ CNRS-FRE 2671\\
Universit\'e de Technologie de Troyes, 12 rue Marie Curie, BP 2060 10010 Troyes, France}}

\newcommand\affildpg{\affiliation{D\'epartement de Photochimie G\'en\'erale\\ CNRS-UMR 7525\\
ENS de Chimie de Mulhouse, 3 rue Alfred Werner, 680993 Mulhouse
cedex, France}}

\newcommand\Ox{O$_2$\xspace}
\newcommand\Eacc{$E_{acc}$\xspace}
\newcommand\Csmax{$C_{smax}$\xspace}
\newcommand\CsmaxD{$C_{smax2}$\xspace}
\newcommand\Cs{$C_{s}$\xspace}
\newcommand\Ddiff{$D_{diff}$\xspace}
\newcommand{\DdiffD}{$D_{diff2}$\xspace}

\title{Modelling the growth of a polymer micro-tip on an optical fiber end}

\author{Malik Hocine}
\affilmops
\author{Nicolas Fressengeas}%
\affilmops
\author{Godefroy Kugel}
\affilmops
\author{Christiane Carr\'e}
\affildpg
\author{Daniel Joseph Lougnot}
\affildpg
\author{Renaud Bachelot}
\affillnio
\author{Pascal Royer}
\affillnio
\date{\today}
\begin{abstract}
Manufacturing end-of-fiber optical components able to realize optical functions ranging from a simple lens to more complex functions such as mode selective components is a decisive but \emph{a priori} complex task owing to the fiber core dimensions. Effective low cost methods allowing to grow polymer components by free-radical photopolymerization using the light coming out of the fiber have recently been reported. A novel phenomenological model of the photopolymerization process underlying is here given and used to simulate the polymer component growth in a three dimensional time-resolved manner. The simulations results are thus used to understand and optimize the component growth conditions, focusing particularly on the role of oxygen either present in the atmosphere or dissolved in the solution.
\\
\end{abstract}

\maketitle
\section{Introduction}
During the past decades, numerous methods for
integrating microlenses on optical fibers (MOFs)
have been reviewed. MOFs have been
manufactured by various methods such as electric arc
melting\cite{kat73jap2756} laser micromachining\cite{pre90ao2692}, chemical etching\cite{eis82ao3470} and photolithography\cite{coh74ao89,bea80,lee85ao3134}. Nevertheless, these procedures are time and
energy consuming and imply stringent control of the
experimental parameters.

A novel low cost manufacturing technique based on free-radical photopolymerization\cite{allen89} was recently reported\cite{bac01,hoc01sm,bac04ol1971,OLEmupo}. It relies on photopolymerization within a photosensitive droplet attached to a cleaved fiber end thanks to surface tension. A beam of visible green light introduced at the other end of the fiber emerges beneath the droplet and thus initiates polymerization.

As recent works show\cite{mon99,mon01jmo191,dor02ol1782}, light propagating in a photosensitive solution in which the refractive index of the polymer is higher than that of the initial formulation can experience self-focusing and propagate in a spatial soliton manner as it is the case in other materials such as Kerr\cite{bar85oc201} or photorefractive media\cite{seg92prl923,fre96pre6866}. Therefore, since the polymer component remains firmly attached to the optical fiber thanks to \emph{hydrogen bonds}, it can be inferred that the component grown following this procedure will be an extension of the fiber core and that its optical function will be governed by its index and shape. In the following, we will thus refer to it as \emph{polymer tip}.

In the following, a thorough 3D time resolved phenomenological model of this component growth on a cleaved fiber end is proposed. Starting with a review of the experimental procedure, an abstract model able to describe the local polymerization rate will be presented, eventually leading to the growth modelling thanks to the coupling with a standard Beam Propagation Method\cite{fei78,Agr89,Cmeos}. It is thus followed by the analysis of the polymer tip growth conditions through simulation in order to understand and master the tip shape and optical function. The model presented in this paper is based on previously published preliminary work \cite{bac01,hoc01sm} and extends the previous models so as to allow their quantitative testing again experimental data.
\section{Experimental description}
\subsection{Photosensitivive solution}
\subsubsection{Composition}
The photopolymerizable solution used in this work  was introduced in previous papers \cite{eco98am411,esp99jpsa2075}. It is made of three basic components: a sensitizer dye, an amine cosynergist, and a multifunctional acrylate monomer, pentaerythritoltriacrylate (PETIA) used as received from the supplier. After irradiation, this component forms the backbone of the polymer network. The cosynergist was methyldiethanolamine (MDEA), while eosin Y (2',4',5',7'-tetrabromo-fluorescein disodium salt) was used as sensitizer dye. This system was developed mainly because of its high sensitivity in the spectral region going from 450 to 550 nm (maximum at 530 nm). In particular, this formulation is specially suited to be irradiated by an argon ion laser (514 or 488 nm), a frequency doubled Nd/YAG laser (532 nm) or a green He-Ne laser at 543.5 nm which we have used in the experiments described in this paper. In addition, this liquid system is very flexible, since it is possible to modify  the nature or the proportions of the different components in order to adjust the physical and chemical properties of the formulation (viscosity, polymerization threshold energy, etc.). Results reported in this paper were obtained with mixtures containing 3 wt.\% eosin and 8 wt.\% methyldiethanolamine.

\subsubsection{Photopolymerization reaction}
After absorption of actinic light by eosin, the triplet state of the dye reacts with the amine to form radicals. Radicals initiate the polymerization of the monomer. Owing to the monomer multifunctionality, the polymer quickly develops into a 3D network. The nonpolymerized formulation stays liquid and can be easily washed out with solvent, such as ethanol, leaving the polymerized part as a quasi-homogeneous transparent polymer component.

However, as previous studies of this photopolymerization process have shown\cite{esp99ass87}, oxygen dissolved within the formulation plays an inhibiting role in the photo-initiation process. As we will see in the following, the oxygen inhibition activity plays a determinant role in the polymer tip growth and thus largely governs its shape and optical function.
\subsection{Experimental set-up}

\begin{figure}
\includegraphics[width=\linewidth]{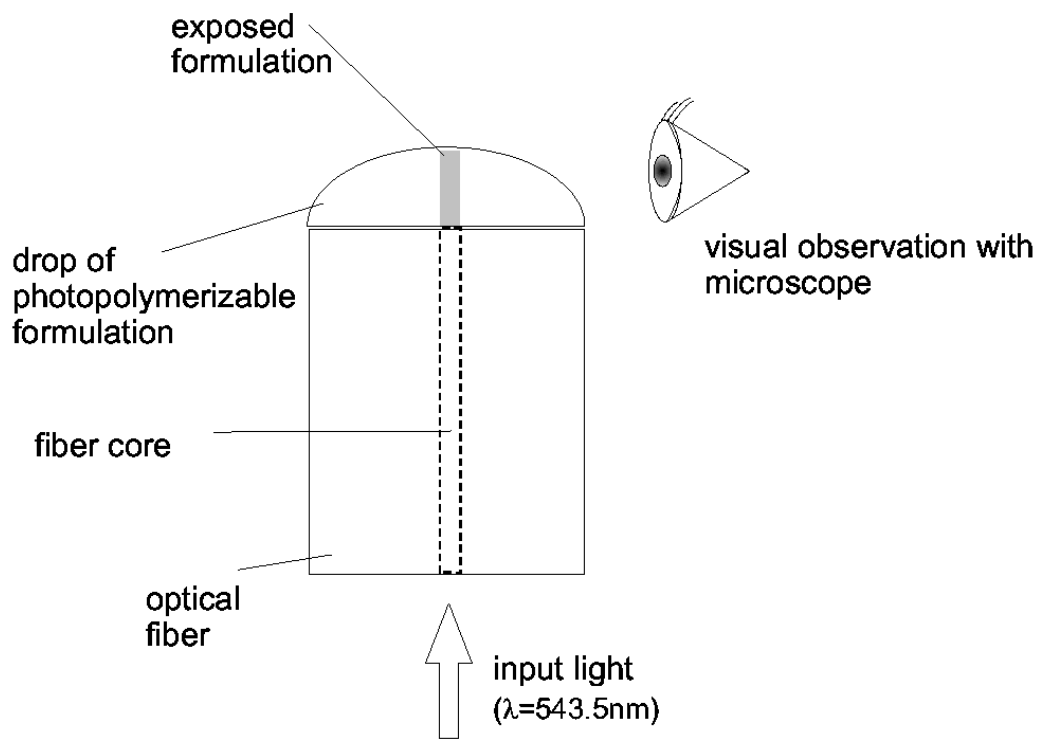}
\caption{\label{experience} Experimental setup: He-Ne 543.5nm laser light emerges from the fiber into the surface tension driven photosensitive droplet, thus initiating the polymerization reaction.}
\end{figure}

As is depicted in Fig.\vref{experience}, the experimental bench set up for the growth of the polymer tip is fairly simple and thus quite low cost. It basically consists in dipping a cleaved optical fiber in a photosensitive solution such as the one described above so as to obtain a hemispherical droplet attached to the end of the fiber by mere surface tension. Coupling a green laser light (for instance a He-Ne laser at a wavelength of 543.5 nm) to the other end of the fiber allows light to emerge below the droplet and initiate polymerization.

The rest of the process consists in adjusting laser power to the desired value and in letting the polymerization occur during the desired exposure time. The resulting polymer tip is still embedded in a non-polymerized monomer solution that can be cleansed by ethanol, for instance, thus leaving the polymer tip standing alone firmly attached to the fiber end.

\subsection{Sample results and applications}

An example of the obtained polymer tip is shown on Fig.\vref{example} with a detail shown on Fig.\vref{extremite}. As expected, the tip looks like an extension of the fiber core and has a length roughly equal to the thickness of the surface tension driven droplet in which its growth took place. On the contrary, the curvature radius of its end is significantly smaller than that of the droplet. Indeed, if they were equal, the tip end would, at that scale, seem rigorously flat. This discrepancy will be thoroughly investigated in the following.

Fig.\vref{kougloff} shows the polymer tip grown from a LP$_{21}$ fiber mode. As can be seen, its shape is a real 3D mold of the electromagnetic intensity emerging from the optical fiber. Preliminary experiments show that this particular tip is a mode selective component: if a plane wave is shone onto it, the excited mode within the fiber happens to be the LP$_{21}$ mode which created the tip.

Other applications based either on the tip optical or mechanical properties can be foreseen. As will be published elsewhere, the polymer tip is able to enhance laser diode to fiber coupling in a significant way. Furthermore, the polymer nature of this tip makes it an attractive low cost non breakable tip for Scanning Near Field Optical Microscopy.

\section{Phenomenological model}
\subsection{Photopolymerization rate empirical model}


The first step of the photopolymerization consists in a photosensitization of the dye ---eosin in our case--- which is excited in its triplet state by the incoming light. This reaction is followed by radical formation which consists in an electron transfer from the amine to the dye, followed by proton transfer. The polymerization is thus initiated and can propagate into a 3D polymer network which tends to solidify the solution.

This reaction main pathway has however to compete with two side reactions involving \Ox which thus inhibits the polymerization process. Indeed, \Ox is able to deactivate the dye triplet state, thus impairing polymerization. It can also peroxydate the free radicals and thus rapidly terminate  the polymer chain\cite{esp99ass87}. Furthermore, these two reactions have been found\cite{dec85} to be much more probable than the polymerization process. Therefore, the deactivation reactions will dominate the polymerization process until \Ox has disappeared from the solution.
This is only a very crude model since it does not take into account \Ox diffusion within the solution towards \Ox depleted region. However, this \emph{local} model can be a good start because it can be expressed in terms of \emph{threshold energy}.

The threshold energy is in this case an energy density above which the time integrated optical energy density at a given point in space ---which we will call the \emph{accumulated energy} \Eacc in the following--- has to grow before polymerization can start.
It is directly linked to the concentration of dissolved \Ox in the formulation to which it is roughly proportional.
Indeed, a typical polymerization response curve is shown on Fig.\vref{cinetique} and exhibits the threshold energy $E_s$.
\subsection{Complete 3D model of light propagation}

        In view of this simple model, the simulation of the growth of the polymer tip requires on the one hand the measurement of the response curve as is given in Fig.\vref{cinetique} and on the other hand the calculation of the accumulated energy \Eacc for every point in space within the solution. The latter can be done knowing the electromagnetic field on the fiber end\cite{jeu} and using a standard Beam Propagation Method (BPM) based on Fourier Transforms\cite{nfo,Cmeos,fei78}. The former allows to get access to the values of the threshold energy $E_s$, the maximum index variation $dn$ and to the polymerization rate linked to the derivative of the curve of Fig.\vref{cinetique} with respect to the accumulated energy. Measurement methods can be found in the litterature\cite{esp99ass87,dor03apl2474}. As will be presented in section III, the phenomenological parameters not directly accessible via measurements (such as $C_s$ and $C_{smax}$ of equation \ref{Es_model} ) are going to be deduced by fitting the simulations to experimental results.

The accumulated energy \Eacc is then computed iteratively using the BPM to know the energy density. The polymerization response curve is then used to evaluate the change in the refraction index at every point in space, thus providing data for the next BPM which will correspond to the next step in time.

As the polymerization response curve used in the simulation is the results of experimental measurements, we have introduced an empirical model which allows its description as a function of the threshold energy $E_s$, the polymerization rate immediately after the threshold ---related to $\alpha$--- and the maximum index variation $dn$:
\begin{equation}
\triangle n=dn \times\tanh \left(\frac{ E_{acc} - E_s} { \alpha
E_s}\right) \times\ h \left( E_{acc}-E_s \right)
\label{polymerization rate}
\end{equation}
where $h$ is the Heaviside unit step function.

Let us point out here that we have taken into account the fact that the polymerization efficiency does depend on the instant local intensity by adding a multiplier $\beta$ to the local intensity whose value is taken between $0$ and $1$ corresponding respectively to null efficiency and maximum one. However, at the typical intensities we use ---2,5mW within the fiber core---, the value of $\beta$ is saturated to $1$\cite{esp99ass87}.

After a number of steps corresponding to the desired exposure time, this process yields a 3D matrix describing the refraction index variation in space. The shape of the polymer tip which is obtained after cleansing is then described by the polymer refraction index $n_p$ ( $1.51\le n_p \le 1.52$ after irradiation, the index of the photopolymerizable being around $1.48$ \cite{dor03apl2474}).
The shape of the tip is thus given by the surface corresponding to a refraction index constant value equal to $n_p$. The typical tip shape obtained with this procedure is shown on Fig.\vref{mupoplate}.

The curvature of the hemispherical droplet is not taken into account in the simulation process. This is justified by the fact that the  radius of curvature of the droplet (larger than the 125 microns fiber diameter) is much larger than the simulation window (on the order of fiber core, a few microns). Therefore, the hemispherical droplet is assumed flat in the simulation region. Furthermore, the droplet thickness of about 35 microns in the simulation region has been measured through electron beam microscopy.

One particular aspect of the above described simulation scheme is that it is resolved in both 3D space and time. This means that light density and thus polymerization degree is allowed to be non-constant through time and non-uniform through space. Particularly, Figure 6, as all simulation figures, has been computed while taking into account the inhomogeneous intensity distribution resulting from the propagation of a Gaussian beam.

\subsection{Dye photo-bleaching}
Our model also needs to consider the loss in absorption due to the consumption of absorbing dye that we must take into account in the BPM process. We have taken into account this phenomenon known as photobleaching in a similar way as the polymerization state. Experimental measurements allow to model phenomenologically the absorption represented by $\gamma$, the absorption coefficient, as a function of the accumulated energy \Eacc at every point of the 3D space in the following way\cite{log90jce872}:

\begin{equation}
  \gamma=\frac{1}{d}\ln\left(1+\frac{e^{d \gamma_0}-1}{e^{\frac{E_{acc}}{\kappa }}}\right)
\end{equation}
where $d$ is the  measurement sample thickness (i.e. an experimental constant), $\gamma_0$ the initial absorption coefficient of the solution and $\kappa$ a characteristic energy density determined experimentally.


Our simulations results show that this photobleaching curve has to be determined by a careful experiment as depicted in the litterature\cite{log90jce872} for it greatly influences the polymer tip shape. In our case, the main parameter $\kappa$ has been measured to 12.5W/cm$^2$.

\subsection{Oxygen diffusion within the solution}
 A simple qualitative comparison between Fig.\vref{extremite} and Fig.\vref{mupoplate} reveals a strong discrepancy between the experimental results and the simulation: the former show a rounded tip with a radius of curvature that goes down to the wavelength in particular experimental conditions whereas the latter evidences a flat end polymer tip.

This discrepancy cannot be attributed to the curvature of the solution droplet because it is several orders of magnitude larger so that the polymer tip would indeed look flat if the curvature was only due to surface tension.
Its origin has to be looked for into what the model does not account for: \Ox diffusion into and within the droplet.


    The major source of \Ox is the growth atmosphere which is generally the ambient atmosphere. Diffusion has then to be accounted for from the droplet surface into the solution for a few tenth of microns of length during the exposure time. The model presented here does  not introduce the classical diffusion Fick law to model \Ox diffusion for two reasons. The first one is a question of computational complexity, which would grow too high for a reasonable 3D model to be developed. The second one is that the corresponding equation is too complicated. The spatially inhomogeneous incident intensity distribution in the formulation at the end of the fiber gives rise to a corresponding nonuniform distribution of both the viscosity and the reactives and photoproducts in the droplet. In particular, the inhomogeneities in the oxygen composition of the droplet should be taken into  account, owing to different reactions with transient species and, simultaneously, to mass diffusion.

It is however possible to infer a phenomenological model from the experiments. Considering that the main process is the diffusion of \Ox from an infinite source into a medium with a homogeneous viscosity constant, the \Ox concentration can be expected to follow an exponential law from a given concentration at the droplet surface. The consequence on the above described model is a spatial modulation of the threshold energy $E_s$ following \Ox concentration. This modulation can be described as follows:

\begin{equation}
C_{smax}\times\exp\left(
\frac{z-d}{D_{diff}}\right)
\label{diff1}
\end{equation}
where \Csmax is linked to \Ox atmospheric concentration on the droplet surface and \Ddiff is called the diffusion length. $d$ is the droplet thickness and $z$ the position of the current point in space counted from the fiber end.


The same analysis can be carried out in the direction transverse to the beam propagation direction: when the beam propagates and initiates polymerization, it consumes oxygen. Following Fick's diffusion law, the \Ox concentration gradient between the beam path and outside the beam path leads to an inward \Ox flux towards the center of the beam. This will impair polymerization on the beam wings. We propose to model this the same way with a spatial modulation in the threshold energy as follows:

\begin{equation}
C_{smax2}\times\exp\left(
\frac{r-d_2}{D_{diff2}}\right)
\label{diff2}
\end{equation}
where $C_{smax2}$ is linked to \Ox concentration within the droplet and \DdiffD is a diffusion length --- \emph{a priori} equal to \Ddiff. $d_2$ is the beam width on the order of the core radius  and $r$ the position of the current point in space counted from the beam axis. This phenomenological model can be understood easily if the beam is considered to be homogeneous with a radius equal to $d_2$. However, if one needs to account for the Gaussian shape of the beam, this simple model needs to be adapted to account for the fact that \Ox depletion varies continuously, potential leading to a raise in the threshold energy in the center of the beam by opposition to its wings. We will not account for that in this paper because we believe that the number of parameters arbitrarily introduced will grow too high for any phenomenological model to be  viable. We will furthermore show that this refinement is not needed to explain the experimental measures. Consequently, undertaking this refinement would imply to reject the phenomenological approach and its computational simplicity and to solve directly the classical laws of diffusion.


If we call $C_s$ the constant threshold energy due to dissolved \Ox during the preparation process and taking into account the diffusion diffusion process modeled through expression \ref{diff1} and \vref{diff2}, the threshold energy can be spatially modeled as follows:

\begin{equation}
 E_s=C_s
 +
 C_{smax}\exp\left(
\frac{z-d}{D_{diff}}\right)
+
C_{smax2}\exp\left(
\frac{r-d_2}{D_{diff2}}\right)
\label{Es_model}
\end{equation}

where \Csmax and \CsmaxD are taken to have the dimensions of an energy density (or fluence).

\subsection{Sample results and model validation}

With this phenomenological model of the photopolymerization process, it it is possible to simulate the time-resolved growth of the polymer micro-tip. Figure \vref{growth1} shows the computed growth of a polymer tip from the fundamental mode of a standard monomomode telecom fiber excited at 543.5$\mu$m  , to be compared to figure \vref{example} which exhibits the experimental shape obtained. The beam profile shown on figure \ref{growth1} does    not seem to exhibit self-focusing as was discussed earlier --- save for the last picture corresponding to a 5s exposure time. This is simply due to the diffraction length of the beam emerging from the fiber, which largely surpasses the droplet thickness --- 117 $\mu$m \emph{vs.} 35$\mu$m .

Figure \vref{growth2} shows the same build-up from a beam emerging from a monomode fiber at 543.5$\mu$m  . The diffraction length is in this case half the droplet thickness and the beam can easily be seen self-focusing, even exhibiting an oscillating behavior as predicted earlier\cite{mon01jmo191}.


\section{Growth conditions design}
\subsection{Solution manufacturing and growth atmosphere control}

Figures \ref{csmax} to \vref{ddiff3mum} alongside tables \vref{table1} to \vref{table3} show the simulations results for varying \Cs, \Csmax and \Ddiff for a given solution and exposure time at a constant optical power. These figures allow to evidence the influences of these model parameters. One can read them easily once knowing that \Cs increases in the horizontal direction and \Csmax in the vertical one.

The atmosphere in which the photosensitive solution is prepared associated to its solubility in the solution plays a crucial role in the concentration of the oxygen dissolved homogeneously within the solution, the model parameter accounting for this phenomenon being \Cs . The analysis of the figures allows to show that an increase in the dissolved oxygen concentration leads to a decrease of the size of the tip.  More precisely, it is the diameter of the polymer tip at the point where it is attached to the fiber which is mostly affected. We can then infer that \Cs will play a role in the size of the modes excited in the tip.

Furthermore, the influence of the oxygen diffusing from the atmosphere can also be retrieved from these figures. An increase of  \Csmax is shown to imply a decrease in the radius of curvature of the polymer tip end and thus a drastic change on the optical function provided by the tip. Let us point out here that we evidenced that it is the growth atmosphere oxygen concentration which designs here the polymer tip optical function.

\subsection{Solution viscosity and oxygen diffusion}
However, once again, the oxygen diffusion from the growth atmosphere towards the solution not only depends on the oxygen partial pressure of the atmosphere and the oxygen solubility but also on the solution itself and its ability to let the oxygen diffuse. Our model accounts for this phenomenon through the oxygen diffusion lengths \Ddiff and \DdiffD. A careful analysis of figures \ref{csmax} to \ref{ddiff3mum} and the corresponding tables allows to evidence that an increase of the oxygen diffusion length leads to a decrease in the radius of curvature of the tip end.

Indeed, a rapid glance at the figures would make the reader think exactly the opposite but attention has to be put on the values of \Cs and \Csmax for each simulation. These varying concentrations of oxygen are simply due to the fact that, experimentally, the oxygen diffusing from the atmosphere does diffuse only in a thickness of a few microns and thus does not influence the photopolymerization process near to the fiber end. This leads to the following relationship which has to be verified for our model to be physically realistic: $C_s \gg C_{smax}\exp\left(\frac{-d}{D_{diff}}\right)$. Furthermore, the homogeneously dissolved oxygen does not play any role at all near the interface with the oxygen rich growth atmosphere, which implies $C_{smax} \gg C_s$, thus leading to an overall relation which has to be verified for all our simulations :

\begin{equation}
  C_s \ll C_{smax} \ll C_s\exp\left(\frac{d}{D_{diff}}\right)
  \label{contrainte}
\end{equation}


A variation of \Ddiff thus implies a correlated variation of \Cs and \Csmax. This, of course, does not imply physically that the oxygen diffusion length in the solution has an influence on the oxygen concentration in the atmosphere but rather that only a small range of varying \Ddiff can physically correspond to a given oxygen concentration. Therefore, equation \ref{contrainte} gives us useful hints on which numerical values to use for our model parameters in order for it to describe correctly the experiments.

The process of diffusion of \Ox within the solution from dark regions to the bright ones which are \Ox depleted is taken in account by the couple of parameters \CsmaxD and \DdiffD. Their influence on the tip shape can be inferred and is confirmed by figure\vref{ddiff2}. Raising either \CsmaxD or \DdiffD leads to a global narrowing of the tip by raising the threshold energy on its sides.

\subsection{Curvature radius prediction and model validation}

The model that is proposed in the previous sections is based on experimental data as far as both polymerization as a function of accumulated energy and the photo-bleaching of the dye are concerned. The four simulation parameters accounting for \Ox diffusion into and within the formulation are on the contrary to be determined by iterative simulation and systematic comparison to experimental data.

It is therefore a systematic experimental measurement campaign that has to be undertaken if one wishes to set the quantitative values for the model parameters. The experimental fit has then to be realized on a value that is both significant and quantitatively within reach both experimentally and through simulation. In this view we have chosen to measure in both cases the radius of curvature of the tip end. It is indeed the tip end which will largely govern the optical function achieved.

In our case, the theoretical fit could not be automated because of the large amount of computing time it would have required: on the contrary, we have manually derived a set of parameters that allows a correct fit between theory and experiment, as can be judged by two significant experimental series which are reported  on figures \ref{compT} and \vref{compI}. The former reports the tip end curvature radius as a function of the exposure time whereas the latter does as a function of the beam power that is injected in the fiber.

Indeed, on the same figures are shown the simulation results obtained from the model parameters given in the caption. As can be expected, threshold energy variation due to the diffusion of the oxygen from the growth atmosphere ---uncontrolled in the experiment--- is much larger than that which is due to the diffusion of oxygen within the droplet. Furthermore, the diffusion length found for both this phenomenon are equal. This again is expected since diffusion occurs in the same medium for both phenomenon. Moreover, the diffusion length  found is around 0.1 $\mu$m . This value is coherent with previous results. Indeed, diffusion coefficients of dye molecules in a polymer host were measured via the holographic grating relaxation techniques\cite{wan90jcp2603,her78jcp2725}. A holographic grating was induced by crossing two equal intensity coherent beams issued from a pulsed laser in order to make the measurement of the slow translational diffusion coefficient feasible. After excitation, the dye concentration was gradually smeared out by diffusion. The corresponding optical grating relaxed with a time  $\tau=\frac{\Lambda^2}{4 \pi^2 D}$ characteristic of the diffusion coefficient $D$, $\Lambda$ being the fringe spacing. Taking $10^{-7}cm^2/s$ for the oxygen diffusion coefficient (by use of the Wilke-Chang correlation \cite{reid77}) and $\Lambda= 10 \mu m $\ (i.e. the lateral dimension of the irradiated zone corresponding to a cylindrical symmetry), a diffusion time of 0.03 s is obtained, corresponding to a value lower than the exposure duration (a few seconds). Then, diffusion of oxygen from the non-irradiated part of the droplet towards the bright area where the monomer is converted into polymer is reasonable. A diffusion length of 0.1 $\mu$m  is here self-consistent due to the reaction of oxygen with all the different active species present in the living polymer. The probability of reaction with an initiating radical, a living monomer radical or a monomer radical trapped in the crosslinked network is sufficiently high to obtain finally a diffusion length around 0.1 $\mu$m  during the growth of a polymer micro-tip. 

In this view, we believe that the quantitative model issued from this experimental fit is then able to predict the polymer tip shape for this photosensitive solution and for various experimental conditions.


Therefore, this quantitative model can now be used to explain and interpret experimental data. For instance, both figures \ref{compT} and \vref{compI} show a change of behavior after an input fluence of 10$\mu$J, the former showing a leap in the curve slope, the latter showing a large break in the curve. The reasons for these behaviors is yet unclear but our simulations allow to show clearly that these behavior changes are related to the tip length. Indeed, the change occurs in both cases when the tip end reaches within an oxygen diffusion length from the droplet end: prior to the break, no influence from the growth atmosphere can be seen whereas after the break, it is atmospheric oxygen which rules the tip end shape.
\section{Conclusion}

The phenomenological model proposed in this paper is based on both experimental measurements and adaptive fit of the four parameters accounting for oxygen diffusion. The set of quantitative experimental measurements presented here allows to get the value for the model parameters for the photosensitive solution we used. Of course, using a different solution would imply another determination of the model parameters. However, once the parameters are set, our model allows to predict the polymer tip shape for various incident beam powers and exposure times. Once this base set, future work sould allow to establish a direct link between our phenomenological parameters and the  physical and chemical parameters characterizing the photopolymerization.

We have not, at this point yet, predicted the optical function of the polymer tip. However, we can assume that the polymer tip, once cleansed with ethanol, shows an homogeneous refraction index equal to that of the polymer at its maximum of polymerization. Indeed, the cleansing will remove  the polymer which has a refraction index lower than the polymer index $n_p=1.51$. The ambient light will then allow to reduce the slight index variations which may be left in the tip by simply allowing the completion of polymerization.

In conclusion, our model allows the prediction off the shape of an optical object which shows an homogeneous refraction index in which electromagnetic waves propagate linearly. We can thus expect to find some of the shelf software able to do the simulation. Figure \vref{pointe} shows the results obtained for a simplified tip by the software FIMMPROP-3D from the society Photon Design which we thank by the way for the figure.

The combination of our model and FIMMPROP-3D thus provides the tool to predict the tip optical function and thus allows the micro-optical manufacturer to grow end-of-fiber polymer micro-components at ultra low cost.

The authors wish to thank the CNRS and Lorraine Region for their partial support of this work.


\begin{table}\center
\begin{tabular}{|c|c|c|c|c|c|}
  \hline Figure & $C_s (J/m^2 )$ & $C_{smax}(J/m^2 )$ & $D_{diff}$($\mu$m) & $\alpha$ & Power($\mu$w) \\\hline
 a & 10$^{3}$ & 10$^{4}$ & 1 & 10& 2.5 \\
 b & 10$^{3}$ & 10$^{8}$ & 1 & 10& 2.5 \\
  c & 10$^{3}$ & 10$^{12}$& 1 & 10 & 2.5 \\
   d & 10$^{3}$ & 10$^{16}$& 1 & 10 & 2.5 \\
  e & 10$^{3}$ & 10$^{18}$& 1 & 10 & 2.5 \\ \hline
 f & 10$^{4}$ & 10$^{5}$ & 1 & 10& 2.5 \\
 g & 10$^{4}$ & 10$^{9}$ & 1 & 10& 2.5 \\
  h & 10$^{4}$ & 10$^{13}$& 1 & 10 & 2.5 \\
   i & 10$^{4}$ & 10$^{17}$& 1 & 10 & 2.5 \\
  j & 10$^{4}$ & 10$^{19}$& 1 & 10 & 2.5 \\ \hline
  k & 10$^{5}$ & 10$^{6}$ & 1 & 10& 2.5 \\
 l & 10$^{5}$ & 10$^{10}$ & 1 & 10& 2.5 \\
  m & 10$^{5}$ & 10$^{14}$& 1 & 10 & 2.5 \\
   n & 10$^{5}$ & 10$^{18}$& 1 & 10 & 2.5 \\
  o & 10$^{5}$ & 10$^{20}$& 1 & 10 & 2.5 \\ \hline
\end{tabular}
\caption{\label{table1} Simulation values for Fig.\vref{csmax}. All simulations are done with an exposure time corresponding to 1s.}
\end{table}

\begin{table}\center
\begin{tabular}{|c|c|c|c|c|c|}
  \hline Figure & $C_s(J/m^2 )$ & $C_{smax}(J/m^2 )$ & $D_{diff}$($\mu$m) & $\alpha$ & Power($\mu$w) \\\hline
 a & 10$^{3}$ & 10$^{4}$ & 2 & 10& 2.5 \\
 b & 10$^{3}$ & 10$^{6}$ & 2 & 10& 2.5 \\
  c & 10$^{3}$ & 10$^{8}$& 2 & 10 & 2.5 \\
   d & 10$^{3}$ & 10$^{9}$& 2 & 10 & 2.5 \\
  e & 10$^{3}$ & 10$^{10}$& 2 & 10 & 2.5 \\ \hline
f & 10$^{4}$ & 10$^{5}$ & 2 & 10& 2.5 \\
 g & 10$^{4}$ & 10$^{7}$ & 2 & 10& 2.5 \\
  h & 10$^{4}$ & 10$^{9}$& 2 & 10 & 2.5 \\
   i & 10$^{4}$ & 10$^{10}$& 2 & 10 & 2.5 \\
  j & 10$^{4}$ & 10$^{11}$& 2 & 10 & 2.5 \\ \hline
 k & 10$^{5}$ & 10$^{6}$ & 2 & 10& 2.5 \\
 l & 10$^{5}$ & 10$^{8}$ & 2 & 10& 2.5 \\
  m & 10$^{5}$ & 10$^{10}$& 2 & 10 & 2.5 \\
   n & 10$^{5}$ & 10$^{11}$& 2 & 10 & 2.5 \\
  o & 10$^{5}$ & 10$^{12}$& 2 & 10 & 2.5 \\ \hline
\end{tabular}–
\caption{\label{table2} Simulation values for Fig.\vref{ddiff2mum}. All simulations are done with an exposure time corresponding to 1s.}
\end{table}

\begin{table}\center
\begin{tabular}{|c|c|c|c|c|c|}
  \hline Figure & $C_s(J/m^2 )$ & $C_{smax}(J/m^2 )$ & $D_{diff}$($\mu$m) & $\alpha$ & Power($\mu$w) \\\hline
  a & 10$^{3}$ & 10$^{4}$ & 3 & 10& 2.5 \\
 b & 10$^{3}$ & 10$^{5}$ & 3 & 10& 2.5 \\
  c & 10$^{3}$ & 10$^{6}$& 3 & 10 & 2.5 \\
   d & 10$^{3}$ & 10$^{7}$& 3 & 10 & 2.5 \\
  e & 10$^{3}$ & 10$^{8}$& 3 & 10 & 2.5 \\ \hline
 f & 10$^{4}$ & 10$^{5}$ & 3 & 10& 2.5 \\
 g & 10$^{4}$ & 10$^{6}$ & 3 & 10& 2.5 \\
  h & 10$^{4}$ & 10$^{7}$& 3 & 10 & 2.5 \\
   i & 10$^{4}$ & 10$^{8}$& 3 & 10 & 2.5 \\
  j & 10$^{4}$ & 10$^{9}$& 3 & 10 & 2.5 \\ \hline
 k & 10$^{5}$ & 10$^{6}$ & 3 & 10& 2.5 \\
 l & 10$^{5}$ & 10$^{7}$ & 3 & 10& 2.5 \\
 m & 10$^{5}$ & 10$^{8}$& 3 & 10 & 2.5 \\
   n & 10$^{5}$ & 10$^{9}$& 3 & 10 & 2.5 \\
  o & 10$^{5}$ & 10$^{10}$& 3 & 10 & 2.5 \\ \hline
    \end{tabular}
\caption{\label{table3} Simulation values for Fig.\vref{ddiff3mum}. All simulations are done with an exposure time corresponding to 1s.}
\end{table}

\clearpage
\section*{List of figure captions}
\begin{figure}
\includegraphics[width=\linewidth]{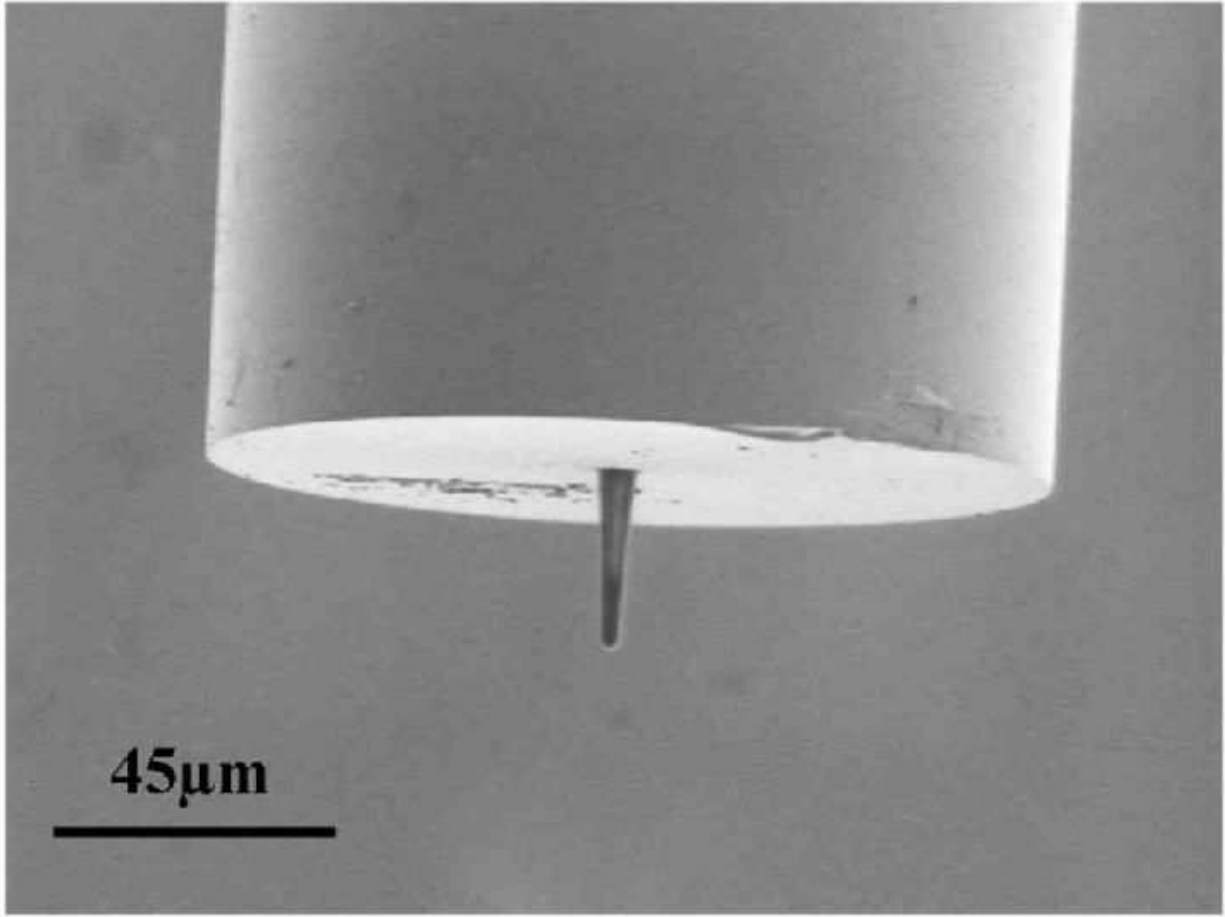}
\caption{\label{example} Electron micrograph of a 30 $\mu$m  long polymer tip grown on the end of monomode fiber  at 543.5nm. Its basis width is on the order of the fiber core: 3 microns.}
\end{figure}

\begin{figure}
\includegraphics[width=\linewidth]{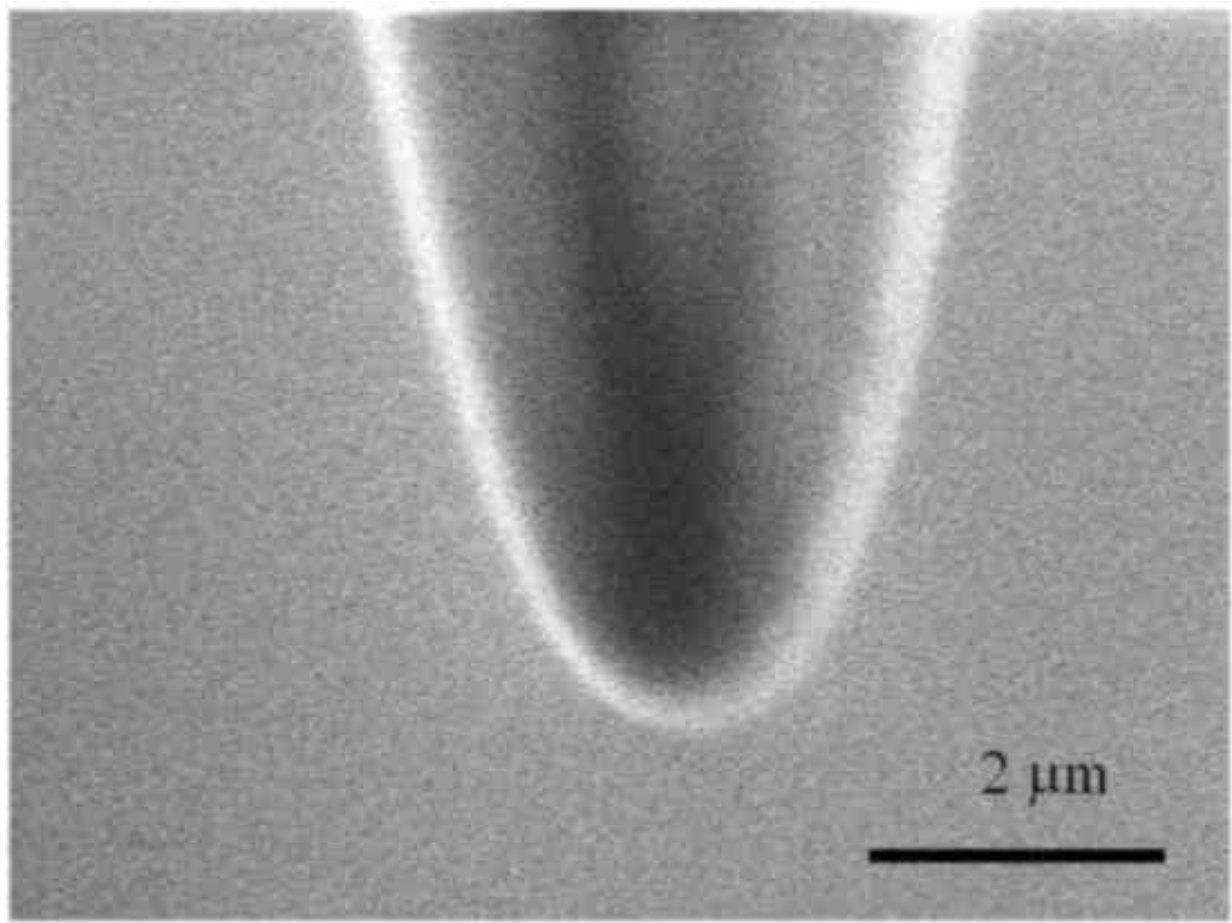}
\caption{\label{extremite} Detail of the end of the polymer tip shown of Fig.\vref{example}.}
\end{figure}

\begin{figure}
\includegraphics[width=\linewidth]{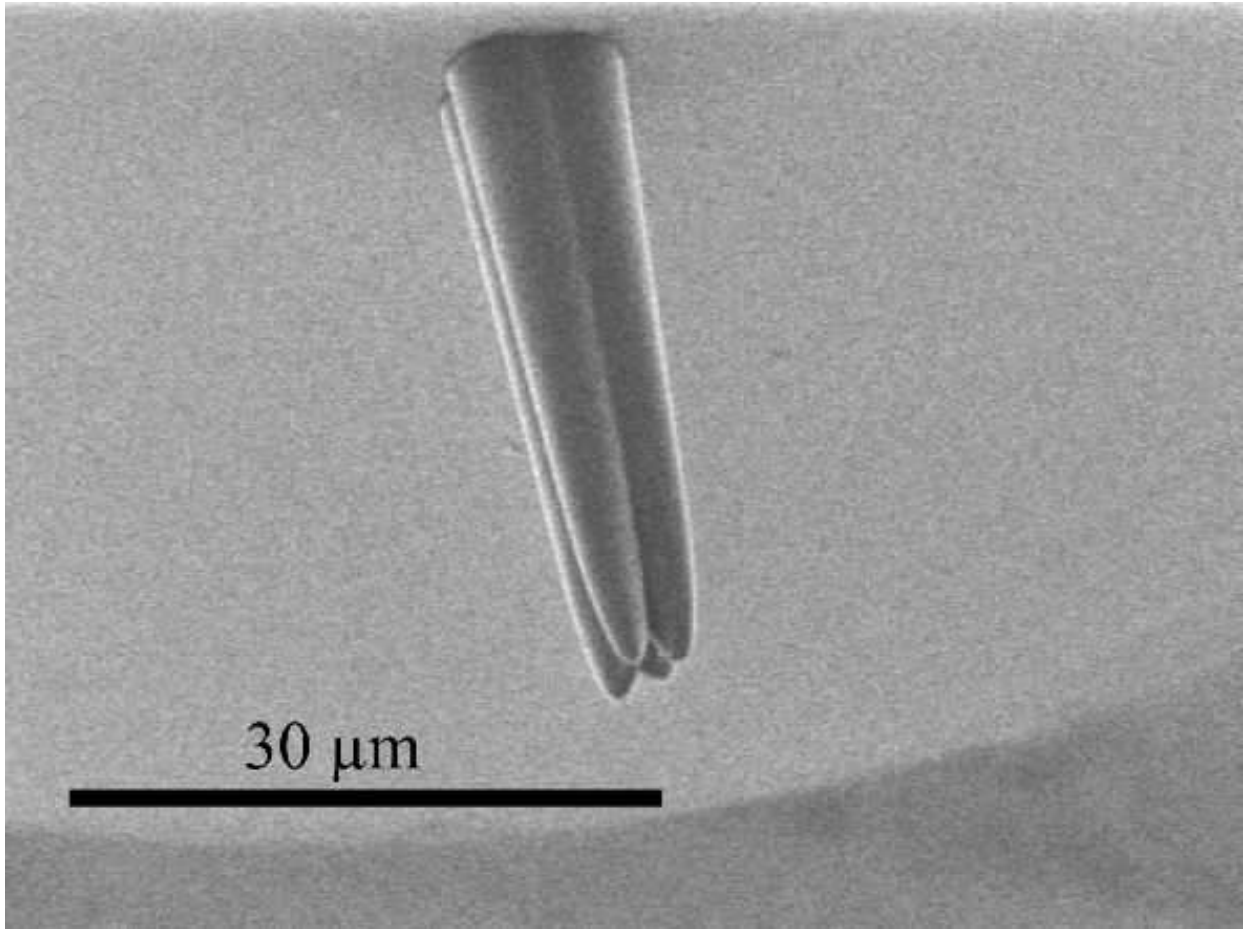}
\caption{\label{kougloff} Polymer tip grown with a LP$_{21}$ mode on the end of a standard telecommunication fiber.}
\end{figure}

\begin{figure}
\includegraphics[width=\linewidth]{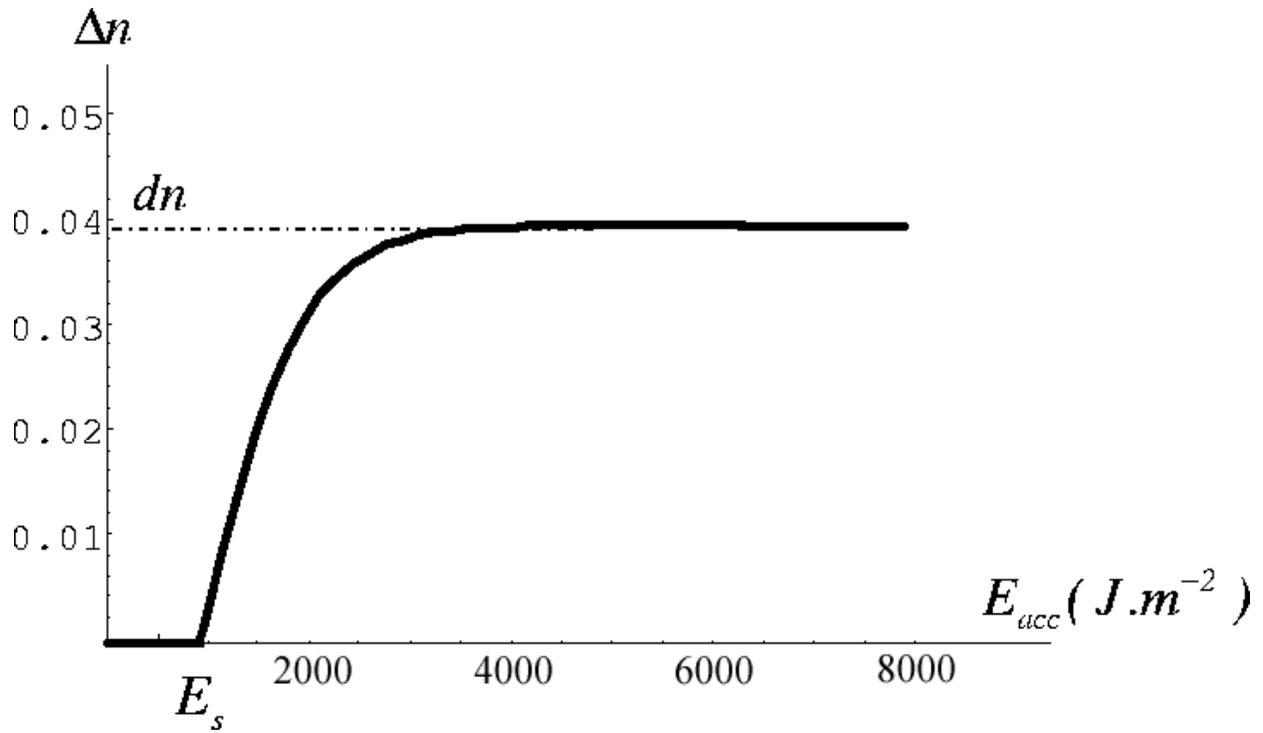}
\caption{\label{cinetique} Typical polymerization curve expressed in terms of index of refraction variation as a function of the accumulated energy \Eacc, the maximum index variation between the monomer and the polymer being $dn=0.04$. The existence of a threshold energy $E_s$ below which no polymerization can occur is evidenced.}
\end{figure}

\begin{figure}
\includegraphics[width=\linewidth]{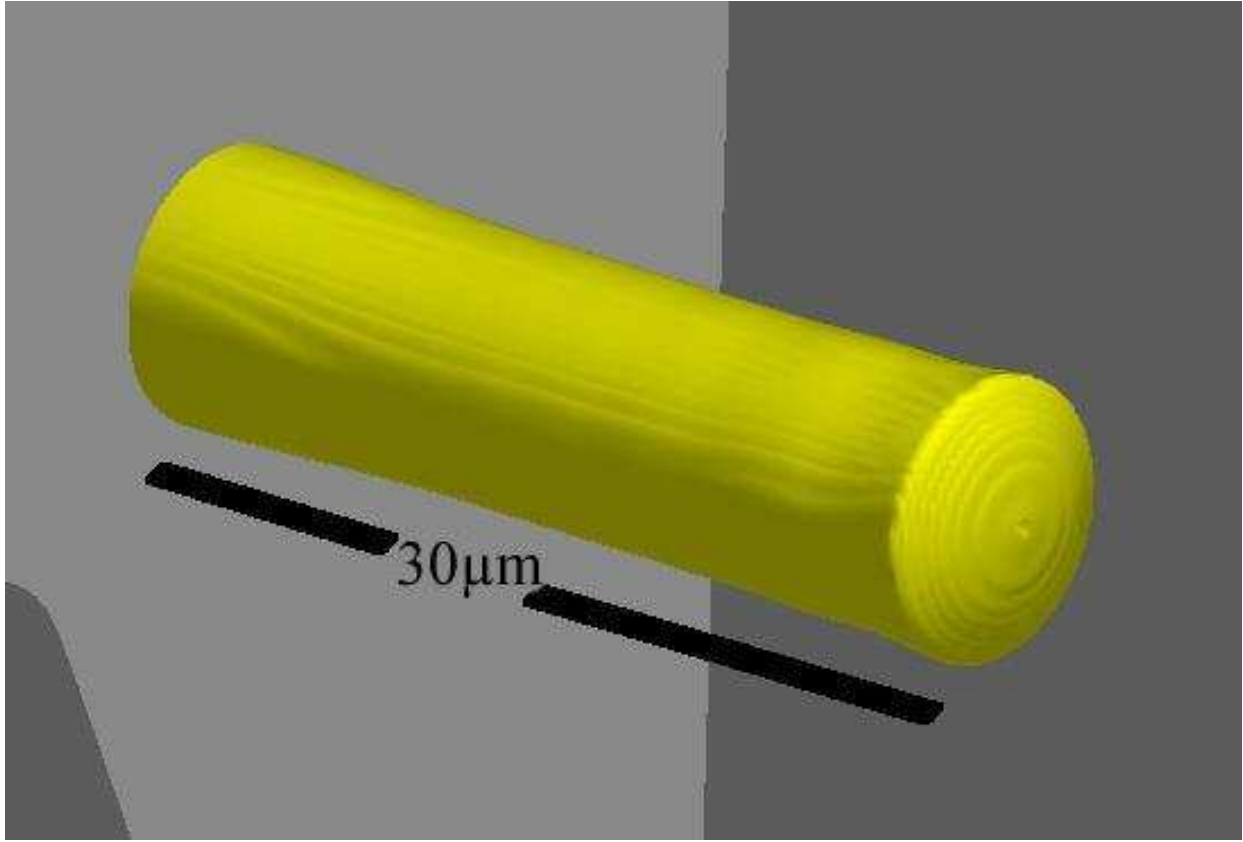}
\caption{\label{mupoplate} Typical shape of a 30$\mu$m  long polymer tip computed using standard BPM and the accumulated energy model shown on Fig.\vref{cinetique} without taking into account \Ox diffusion into the solution. 
}
\end{figure}

\begin{figure*}
\includegraphics[width=\linewidth]{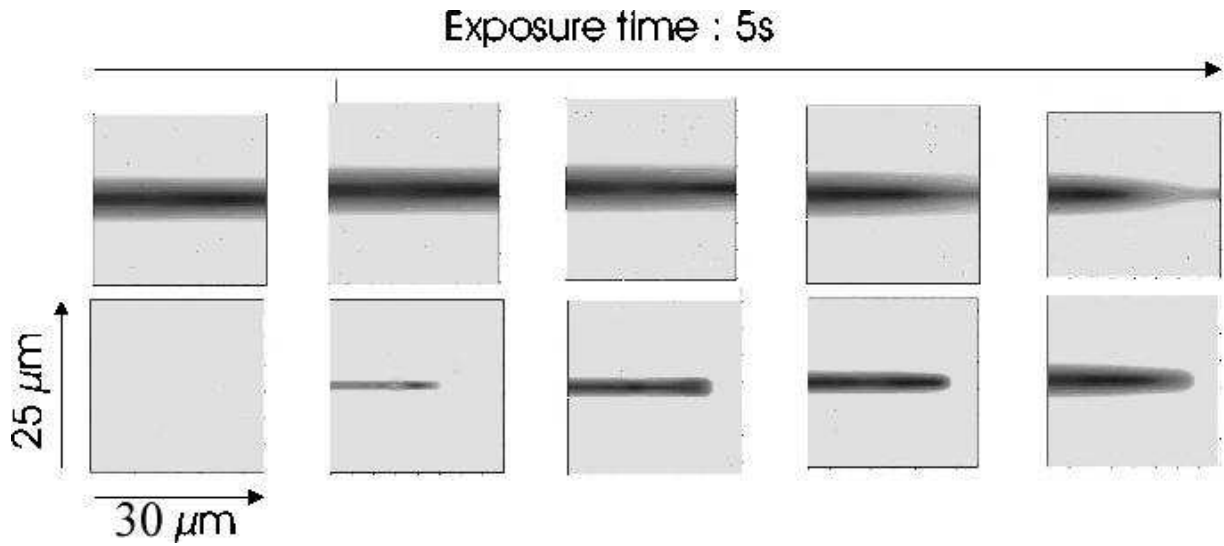}
\caption{\label{growth1} Simulated growth of a polymer tip on the cleaved end of monomode telecom 1.55$\mu$m  fiber excited in 543.5$\mu$m  in its fundamental mode (bottom), alongside the beam profile during the growth (top). The beam emerges from the fiber on the left.}
\end{figure*}

\begin{figure*}
\includegraphics[width=\linewidth]{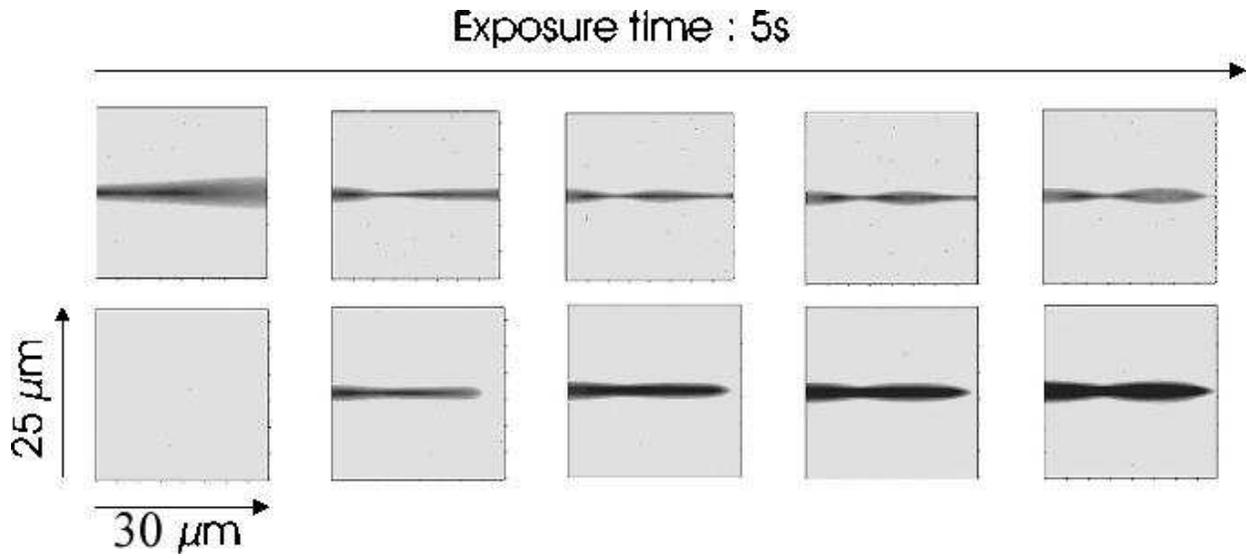}
\caption{\label{growth2} Simulated growth of a polymer tip on the cleaved end of monomode 543.5nm fiber (bottom), alongside the beam profile during the growth (top). The beam emerges from the fiber on the left.}
\end{figure*}

\begin{figure*}
\includegraphics[width=\linewidth]{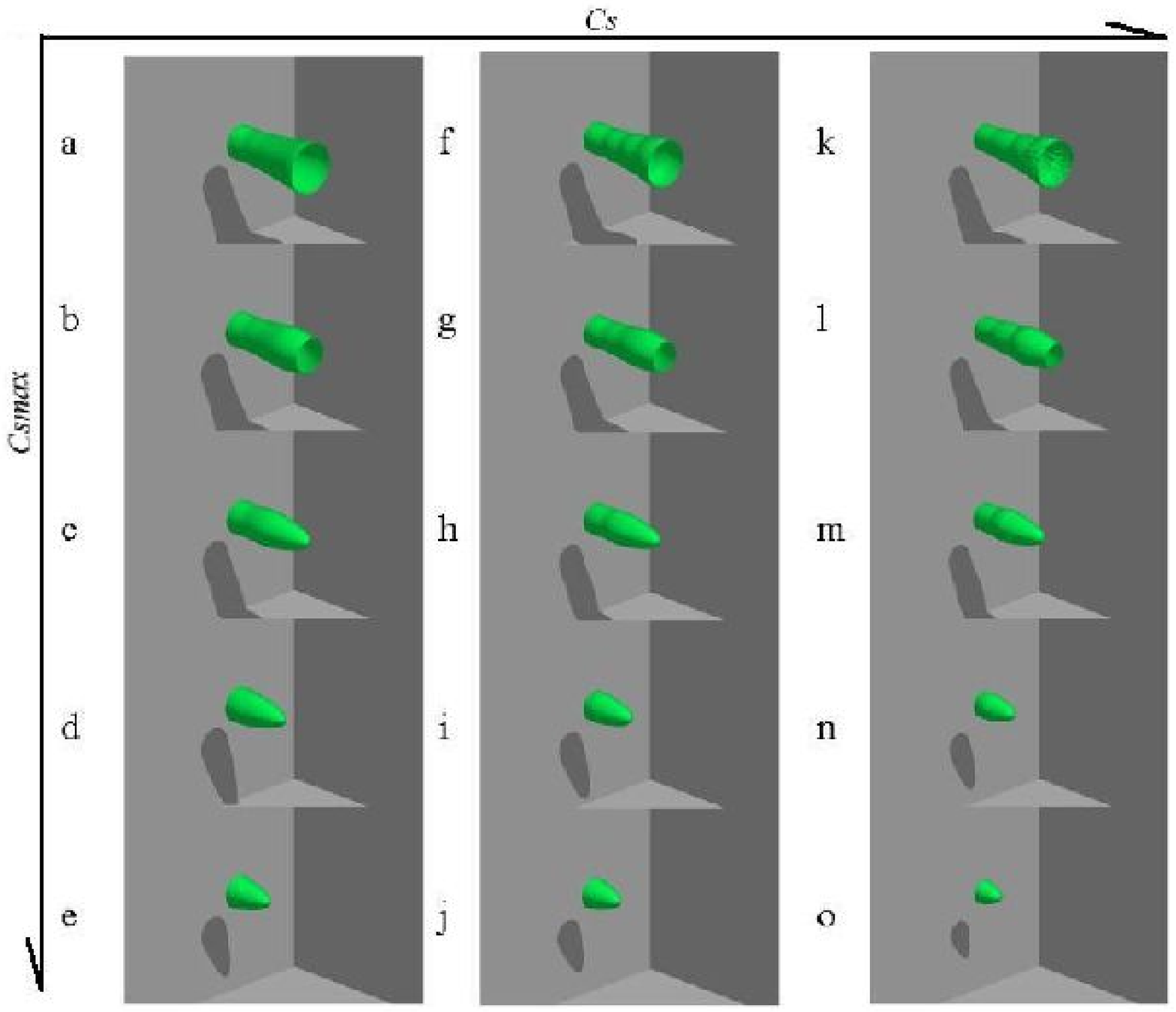}
\caption{\label{csmax}Simulated influence of dissolved oxygen ($C_s$) and atmosphere \Ox concentration (\Csmax) for a diffusion length $D_{diff} = 1\mu m $. From (a) to (e) $C_s = 10^3J/m^2 $ and \Csmax rises. From  (f) to (j) $C_s = 10^4J/m^2 $ and \Csmax rises. From (k) to (o) $Cs = 10^5J/m^2 $ and \Csmax rises. Precise values are given in Table \vref{table1}. The fiber-end to which is attached the tip is located on the left-hand side. The hollow ends of tips that can be seen on figures \ref{csmax} through \ref{ddiff3mum} are a rendering artifact and mean that the tips are actually flat.}
\end{figure*}

\begin{figure*}
\includegraphics[width=\linewidth]{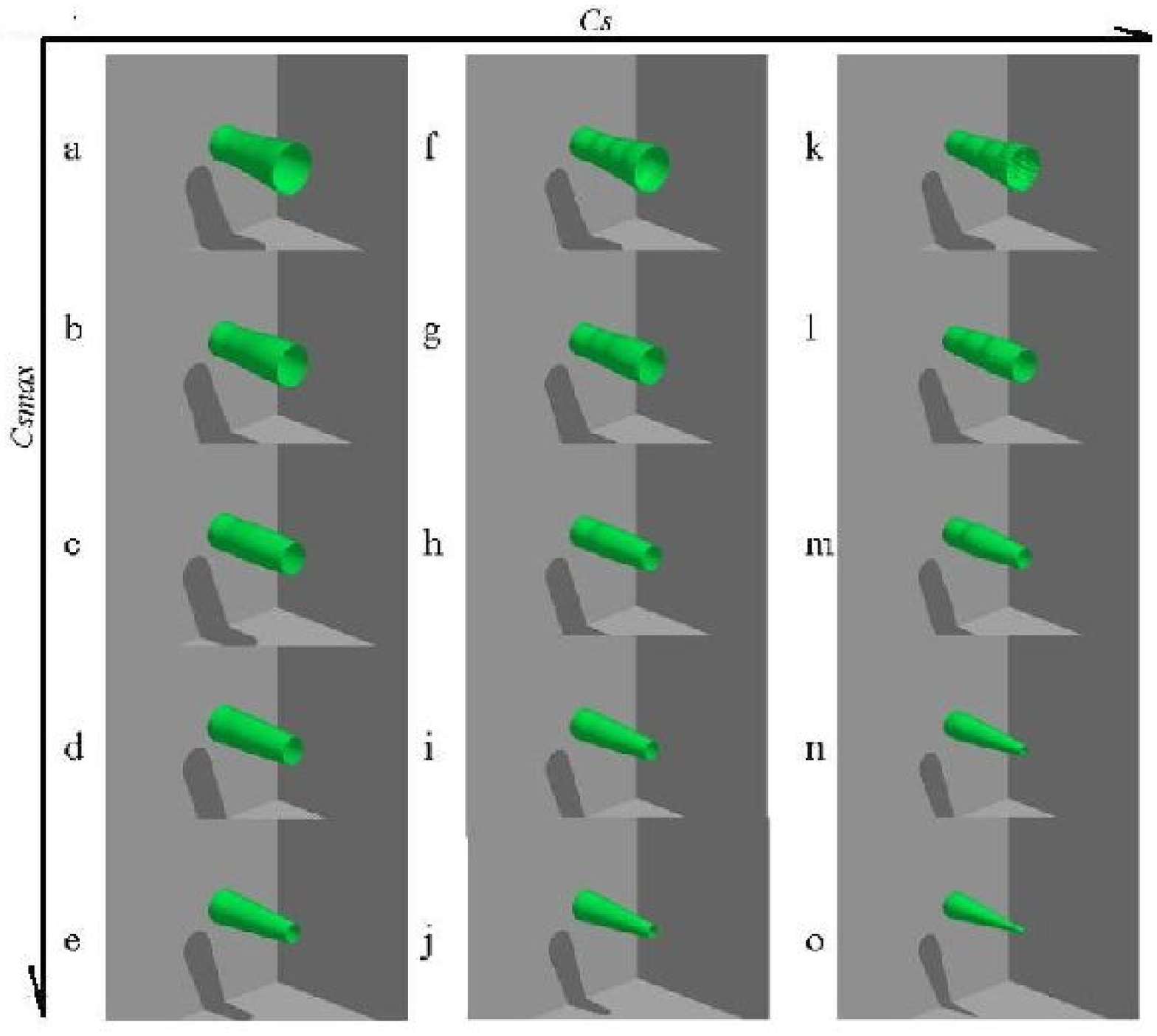}
\caption{\label{ddiff2mum}Simulated influence of dissolved oxygen ($C_s$) and atmosphere \Ox concentration (\Csmax) for a diffusion length $D_{diff} = 2\mu m $. From (a) to (e) $C_s = 10^3 J/m^2$ and \Csmax rises. From  (f) to (j) $C_s = 10^4J/m^2 $ and \Csmax rises. From (k) to (o) $Cs = 10^5J/m^2 $ and \Csmax rises. Precise values are given in Table \vref{table2}.  The fiber-end to which is attached the tip is located on the left-hand side.}
\end{figure*}

\begin{figure*}
\includegraphics[width=\linewidth]{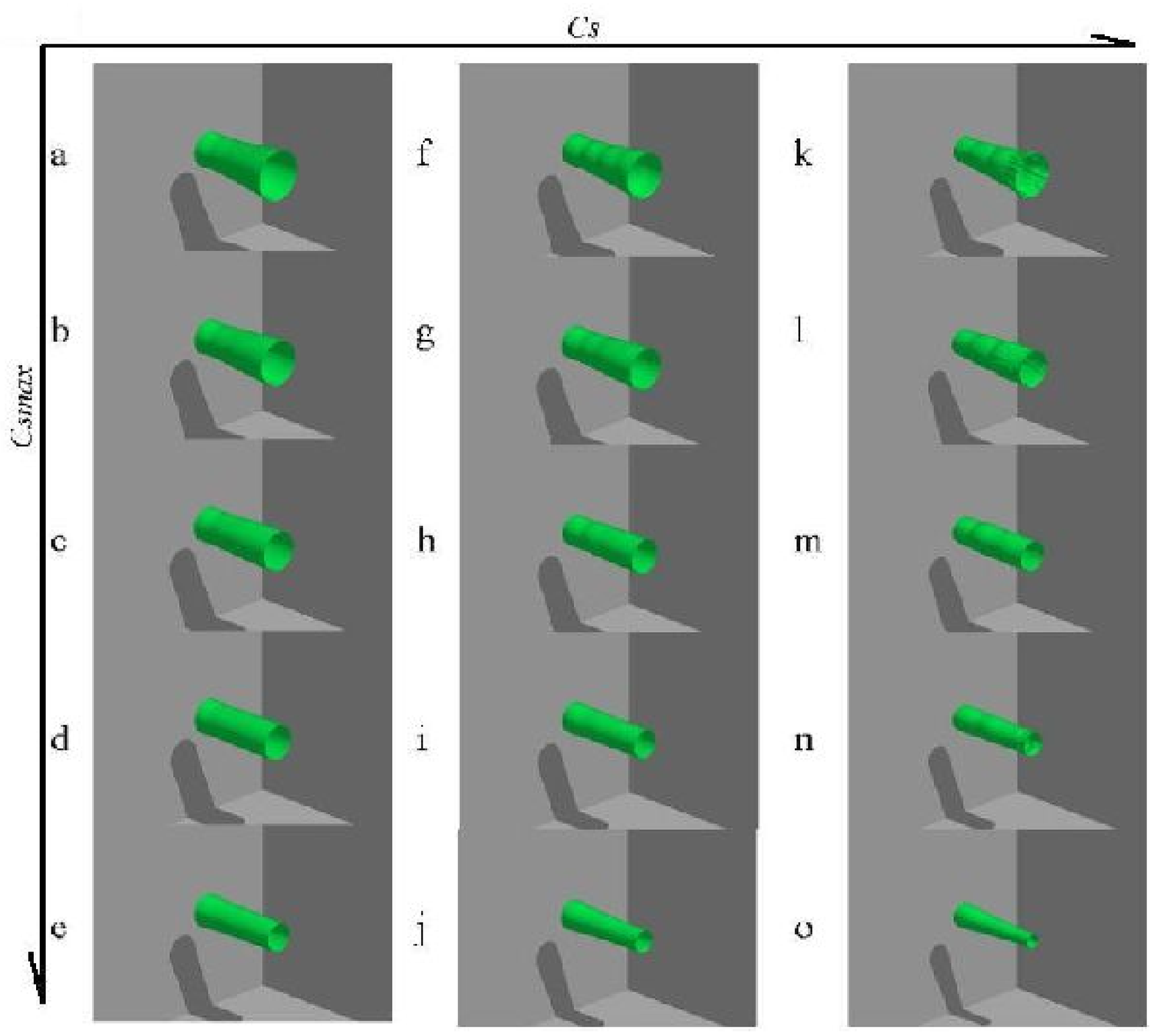}
\caption{\label{ddiff3mum}Simulated influence of dissolved oxygen ($C_s$) and atmosphere \Ox concentration (\Csmax) for a diffusion length $D_{diff} = 3\mu m $. From (a) to (e) $C_s = 10^3J/m^2 $ and \Csmax rises. From  (f) to (j) $C_s = 10^4J/m^2 $ and \Csmax rises. From (k) to (o) $Cs = 10^5J/m^2 $ and \Csmax rises. Precise values are given in Table \vref{table3}.  The fiber-end to which is attached the tip is located on the left-hand side.}
\end{figure*}

\begin{figure}
  \includegraphics[width=\linewidth]{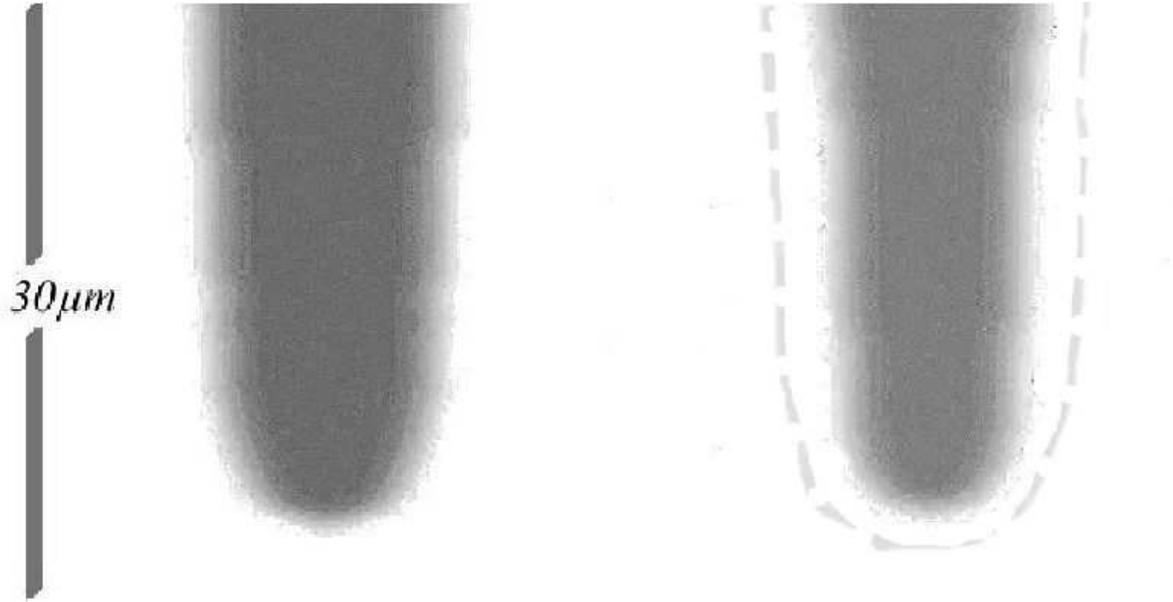}
\caption{Left: simulated polymer tip without taking into account diffusion within the droplet. Right: droplet inner diffusion taken in account. Simulation data: C$_s$=5$\times$10$^2$J/m$^2$, C$_{smax}$=10$^8$J/m$^2$,
\DdiffD=\Ddiff=0.1$\mu$m,
C$_{smax2}$=10$^4$J/m$^2$ and an exposure time of 1s. The dotted line around the right-hand tip is a shadow of the left-hand one meant to ease comparison.\label{ddiff2}}
\end{figure}

\begin{figure}
  \includegraphics[width=\linewidth]{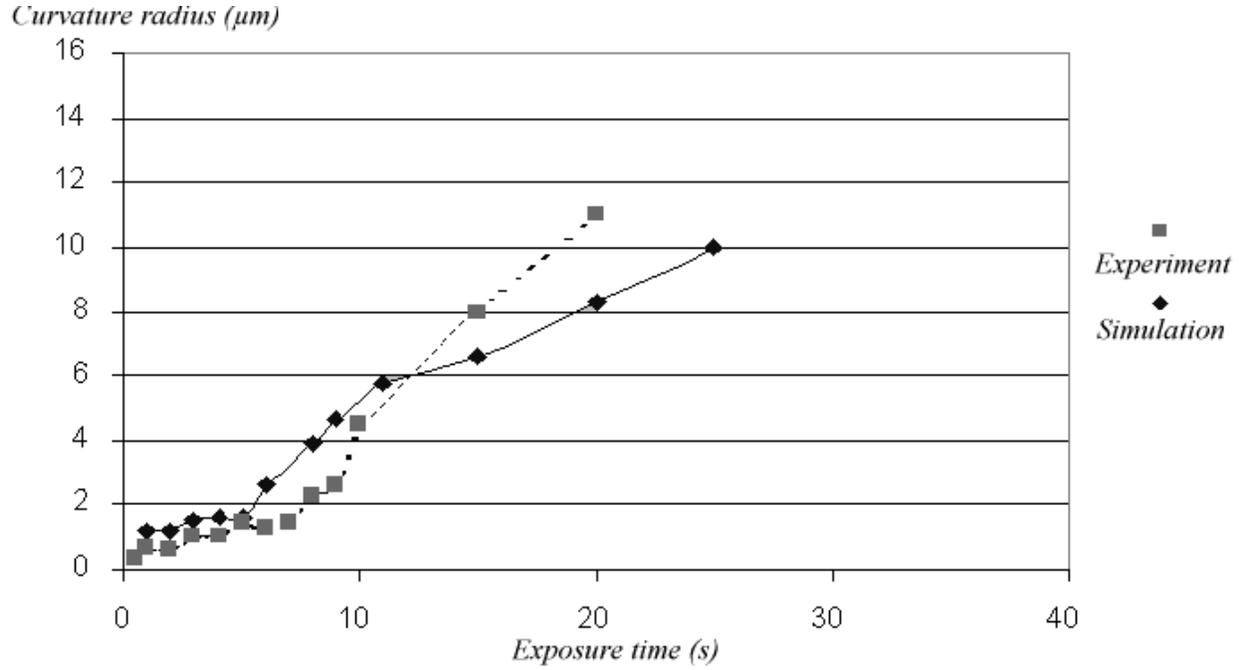}
    \caption{\label{compT} Experimentally measured curvature radii are compared to simulated ones for varying exposure times and for a beam power of 2,5$\mu$W, an experimental eosin concentration of 3\% and the following simulation data: C$_s$=
5$\times$10$^2$J/m$^2$ ,C$_{smax}=$10$^8$J/m$^2$, D$_{diff}$=D$_{diff2}$= 0,1$\mu$m,
C$_{smax2}$=10$^4$J/m$^2$ and $\alpha=10$.}
\end{figure}

\begin{figure}
  \includegraphics[width=\linewidth]{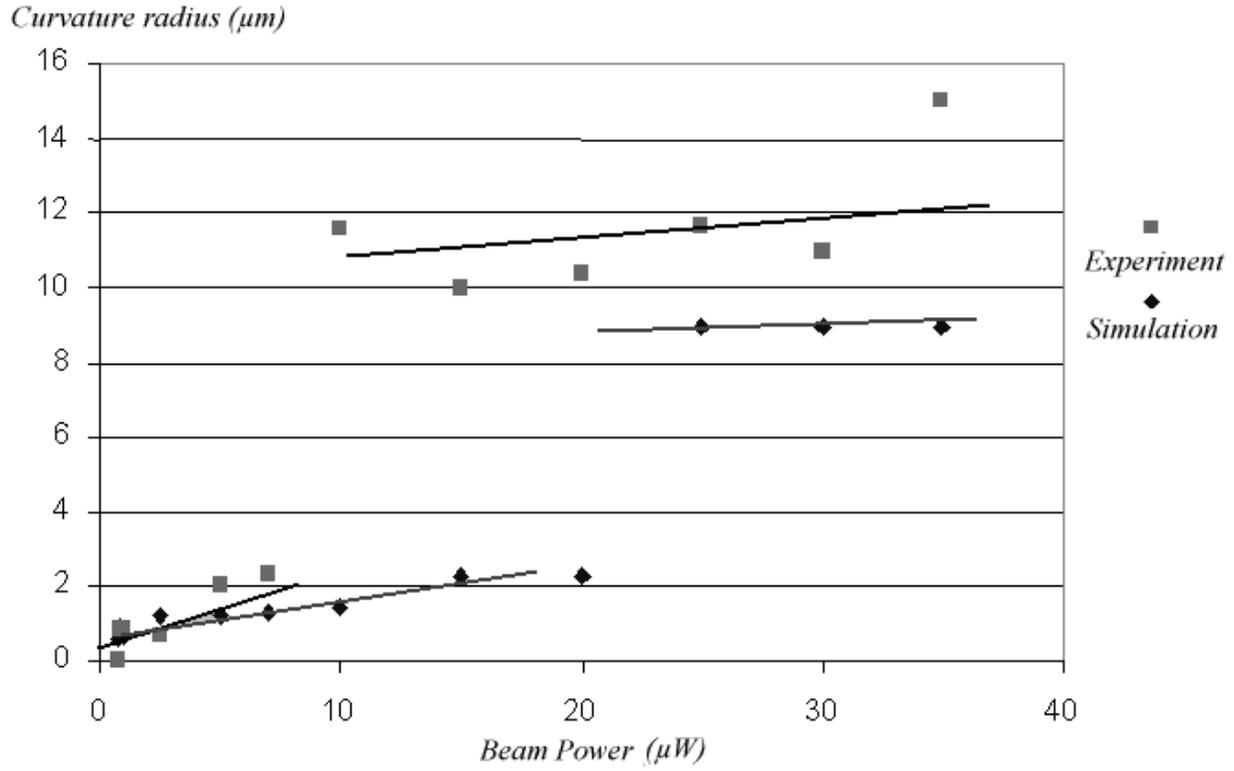}
    \caption{\label{compI} Experimentally measured curvature radii are compared to simulated ones for varying output beam powers and for an exposure time of 1s, an experimental eosin concentration of 3\% and the following simulation data: C$_s$=
5$\times$10$^2$J/m$^2$ ,C$_{smax}=$10$^8$J/m$^2$, D$_{diff}$=D$_{diff2}$= 0,1$\mu$m,
C$_{smax2}$=10$^4$J/m$^2$ and $\alpha=10$. The straight lines are a mere guide to the eyes.}
\end{figure}

\begin{figure}
\includegraphics[width=\linewidth]{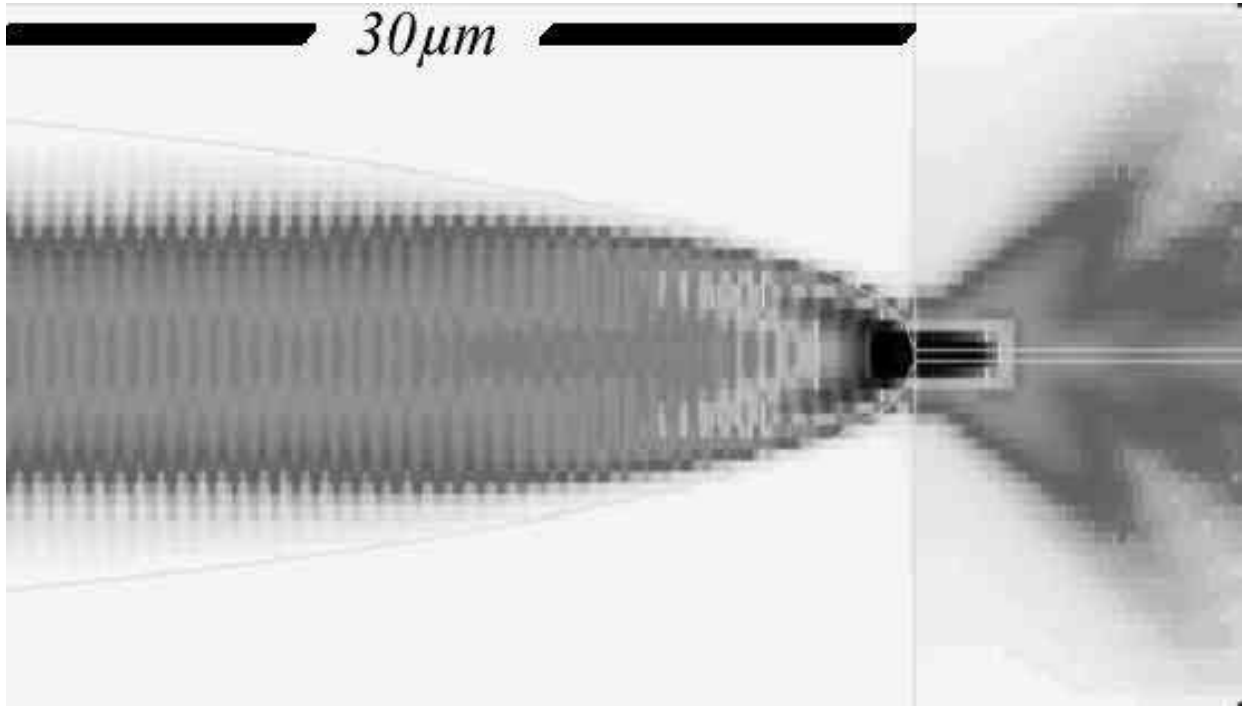}
\caption{\label{pointe}Simulated propagation of light in a sample polymer tip: the initially plane wave undergoes strong focusing prior to reaching the tip end, thus providing high enlargement of the numerical aperture of the system.}
\end{figure}

\begin{enumerate}
\item{\label{fakeexperience} Experimental setup: He-Ne 543.5nm laser light emerges from the fiber into the surface tension driven photosensitive droplet, thus initiating the polymerization reaction.}

\item{\label{fakeexample} Electron micrograph of a 30 $\mu$m  long polymer tip grown on the end of monomode fiber  at 543.5nm. Its basis width is on the order of the fiber core: 3 microns.}

\item{\label{fakeextremite} Detail of the end of the polymer tip shown of Fig.\vref{example}.}

\item{\label{fakekougloff} Polymer tip grown with a LP$_{21}$ mode on the end of a standard telecommunication fiber.}

\item{\label{fakecinetique} Typical polymerization curve expressed in terms of index of refraction variation as a function of the accumulated energy \Eacc, the maximum index variation between the monomer and the polymer being $dn=0.04$. The existence of a threshold energy $E_s$ below which no polymerization can occur is evidenced.}

\item{\label{fakemupoplate} Typical shape of a 30$\mu$m  long polymer tip computed using standard BPM and the accumulated energy model shown on Fig.\vref{cinetique} without taking into account \Ox diffusion into the solution. 
}

\item{\label{fakegrowth1} Simulated growth of a polymer tip on the cleaved end of monomode telecom 1.55$\mu$m  fiber excited in 543.5$\mu$m  in its fundamental mode (bottom), alongside the beam profile during the growth (top). The beam emerges from the fiber on the left.}

\item{\label{fakegrowth2} Simulated growth of a polymer tip on the cleaved end of monomode 543.5nm fiber (bottom), alongside the beam profile during the growth (top). The beam emerges from the fiber on the left.}

\item{\label{fakecsmax}Simulated influence of dissolved oxygen ($C_s$) and atmosphere \Ox concentration (\Csmax) for a diffusion length $D_{diff} = 1\mu m $. From (a) to (e) $C_s = 10^3J/m^2 $ and \Csmax rises. From  (f) to (j) $C_s = 10^4J/m^2 $ and \Csmax rises. From (k) to (o) $Cs = 10^5J/m^2 $ and \Csmax rises. Precise values are given in Table \vref{table1}. The fiber-end to which is attached the tip is located on the left-hand side. The hollow ends of tips that can be seen on figures \ref{csmax} through \ref{ddiff3mum} are a rendering artifact and mean that the tips are actually flat.}

\item{\label{fakeddiff2mum}Simulated influence of dissolved oxygen ($C_s$) and atmosphere \Ox concentration (\Csmax) for a diffusion length $D_{diff} = 2\mu m $. From (a) to (e) $C_s = 10^3 J/m^2$ and \Csmax rises. From  (f) to (j) $C_s = 10^4J/m^2 $ and \Csmax rises. From (k) to (o) $Cs = 10^5J/m^2 $ and \Csmax rises. Precise values are given in Table \vref{table2}.  The fiber-end to which is attached the tip is located on the left-hand side.}

\item{\label{fakeddiff3mum}Simulated influence of dissolved oxygen ($C_s$) and atmosphere \Ox concentration (\Csmax) for a diffusion length $D_{diff} = 3\mu m $. From (a) to (e) $C_s = 10^3J/m^2 $ and \Csmax rises. From  (f) to (j) $C_s = 10^4J/m^2 $ and \Csmax rises. From (k) to (o) $Cs = 10^5J/m^2 $ and \Csmax rises. Precise values are given in Table \vref{table3}.  The fiber-end to which is attached the tip is located on the left-hand side.}

\item{Left: simulated polymer tip without taking into account diffusion within the droplet. Right: droplet inner diffusion taken in account. Simulation data: C$_s$=5$\times$10$^2$J/m$^2$, C$_{smax}$=10$^8$J/m$^2$,
\DdiffD=\Ddiff=0.1$\mu$m,
C$_{smax2}$=10$^4$J/m$^2$ and an exposure time of 1s. The dotted line around the right-hand tip is a shadow of the left-hand one meant to ease comparison.\label{fakeddiff2}}

    \item{\label{fakecompT} Experimentally measured curvature radii are compared to simulated ones for varying exposure times and for a beam power of 2,5$\mu$W, an experimental eosin concentration of 3\% and the following simulation data: C$_s$=
5$\times$10$^2$J/m$^2$ ,C$_{smax}=$10$^8$J/m$^2$, D$_{diff}$=D$_{diff2}$= 0,1$\mu$m,
C$_{smax2}$=10$^4$J/m$^2$ and $\alpha=10$.}

    \item{\label{fakecompI} Experimentally measured curvature radii are compared to simulated ones for varying output beam powers and for an exposure time of 1s, an experimental eosin concentration of 3\% and the following simulation data: C$_s$=
5$\times$10$^2$J/m$^2$ ,C$_{smax}=$10$^8$J/m$^2$, D$_{diff}$=D$_{diff2}$= 0,1$\mu$m,
C$_{smax2}$=10$^4$J/m$^2$ and $\alpha=10$. The straight lines are a mere guide to the eyes.}

\item{\label{fakepointe}Simulated propagation of light in a sample polymer tip: the initially plane wave undergoes strong focusing prior to reaching the tip end, thus providing high enlargement of the numerical aperture of the system.}

\end{enumerate}
\clearpage

–2085

\clearpage

\end{document}